\documentclass{he_symp}
\usepackage{psfig,graphicx,epsfig}
\usepackage{color}
\usepackage{amsmath,amssymb,epic,eepic,array}
\begin{document}
\renewcommand{\FirstPageOfPaper }{ 183}\renewcommand{\LastPageOfPaper }{ 192}
\def\ltsima{$\; \buildrel < \over \sim \;$}
\def\gtsima{$\; \buildrel > \over \sim \;$}
\def\lsim{\lower.5ex\hbox{\ltsima}}
\def\gsim{\lower.5ex\hbox{\gtsima}}
\def\lapp{\ifmmode\stackrel{<}{_{\sim}}\else$\stackrel{<}{_{\sim}}$\fi}
\def\gapp{\ifmmode\stackrel{>}{_{\sim}}\else$\stackrel{<}{_{\sim}}$\fi}
\def\simless{\mathbin{\lower 3pt\hbox{$\rlap{\raise 5pt\hbox{$\char'074$}}\mathchar"7218$}}} 
\def\simgreat{\mathbin{\lower 3pt\hbox{$\rlap{\raise 5pt\hbox{$\char'076$}}\mathchar"7218$}}} 
\def\vion{{v_{\rm{ion}}}}
\def\vrecoil{v_{\rm{NS},\infty}}
\def\vrecB{v_{\rm{B},\infty}}
\def\vion{{v_{\rm{ion}}}}
\def\porb{{P_{\rm{orb}}}}
\def\ain{{a_{\rm{in}}}}
\def\stil{{{\tilde \sigma}}} 
\def\erg{\rm {erg}} \def\X{X_{\rm{MSP}}}
\def\kms{{\rm\,km\,s^{-1}}} 
\def\persec{{\rm s^{-1}}} 
\def\pc3{{\rm pc^{-3}}} 
\def\msun{M_\odot} 
\def\df{\rm {df}} 
\def\hz{{\rm\,Hz\,}} 
\def\mmean{\langle m\rangle } 
\def\vmean{\langle v\rangle } 
\def\v2mean{\langle v^2\rangle } 
\def\mbhbig{M_{\rm {BH}}} 
\def\mbh2{m_{\rm {BH}}} 
\def\mubh{\mu_{\rm {BH}}} 
\def\bhtot{M_{\rm {BH,t}}} 
\def\mdeg{m^*_{\rm{c}}} 
\def\mcom{m_{\rm{c}}}
\def\mpul{m_{\rm{PSR-A}}} 
\def\apul{a_{\rm{PSR-A}}} 
\def\rpul{r_{\rm{PSR-A}}} 
\def\mtot{M_{\rm {BH,tot}}} 
\def\df{\rm {df}} 
\def\be{\begin{equation}} 
\def\ee{\end{equation}}
\def\baray{\begin{eqnarray}}
\def\earay{\end{eqnarray}}

\title{The peculiar millisecond pulsars in the globular clusters 
NGC~6397 and NGC~6752}
\author{Andrea Possenti\inst{1}}  
\institute{Osservatorio Astronomico di Bologna,
Via Ranzani 1, 40127 Bologna, Italy}
\maketitle

\begin{abstract}
The millisecond pulsars discovered in globular clusters are diagnostic
tools for studying the dynamics of the clusters, the intracluster
ionized gas and the evolution of the binaries embedded in the cluster.
In the first two years of a search at 1.4 GHz in progress at the Parkes 
radio telescope 12 new millisecond pulsars have been discovered
in 6 globular clusters in which no pulsars were previously known. The
scientific fall-back of these discoveries is here discussed for the case of
two globular clusters: NGC~6397 and NGC~6752.
\end{abstract}

\section{Introduction}
Millisecond pulsars (MSPs) are old neutron stars {\it recycled}
through transfer of matter and angular momentum from a mass losing
companion in a binary system (e.g. Alpar et al. 1982; Smarr \&
Blandford 1976; Bhattacharya \& van den Heuvel 1991; Kulkarni \&
Anderson 1996).  They are massive and almost pointlike objects,
appearing very good test masses for probing gravitational
effects. Most of them are also extremely stable clocks, allowing very
accurate measurements of their rotational parameters and position in
the sky. Finally they are the endpoints of a complex stellar evolution
in binary systems, involving not yet completely understood mechanisms
like common envelope, accretion disk instabilities, magnetic braking,
irradiation, evaporation.

Besides evolution of a primordial binary, in a Globular Cluster (GC)
another formation channel for the MSPs is available: in fact exchange
interactions in the ultra-dense stellar environment of the cluster
core can sustain the formation of various kinds of binaries suitable
for recycling neutron stars (Davies \& Hansen 1998; Rasio, Pfahl \&
Rappaport 2001).  As a consequence, despite the large difference in
total mass between the disk of the Galaxy and the Globular Cluster
system, about 50\% of the entire MSP population has been found in the
latter.

In particular millisecond pulsars found in globular cluster
have proven to be valuable tools for studying:
\begin{itemize}
\item the potential well of the globular cluster 
(e.g. Phinney 1992; Camilo et al. 2000);
\item the dynamical interactions in the globular cluster core
(e.g. Phinney \& Sigurdsson 1991);
\item  the gas content in the globular cluster 
(e.g. Freire et al. 2001)
\item  the neutron star retention in the globular cluster 
(e.g. Rappaport et al. 2001);
\item  the binary evolution in a crowded environment 
(e.g. Rasio et al. 2000).
\end{itemize}  

Unfortunately, the MSPs in GC are sources difficult to be
detected. The main reason for their elusiveness is that {\it GC-MSPs
are often distant pulsars in close binary systems}.  Their large
distances {\it (i)} make their fluxes typically very small and {\it
(ii)} strongly distort their signals due to the dispersive effects of
propagation through the interstellar medium.  Their inclusion in tight
binaries {\it (iii)} causes Doppler-shift changes of the apparent spin
period and sometimes {\it (iv)} makes the radio signal periodically
obscured by eclipses.

As a consequence, except for the relevant case of 47~Tucanae (Camilo
et al. 2000), the discovery rate of these objects strongly declined in
the second half of the 1990s: in the 7 years from 1987 (when the first
GC-MSP, B1821$-$24, was discovered in M28 at Jodrell Bank: Lyne et
al. 1987) to 1994 (Biggs et al. 1994) 32 GC-MSPs entered the catalog,
whereas no new source was published in the following 5 years.
 
Since two years a new deep search has been running at the Parkes radio
telescope; it involves people from the Cagliari Astronomical
Observatory (N. D'Amico), from the Australia Telescope National
Facility (R.~N. Manchester and J. Sarkissian), from the Jodrell Bank
Observatory (A.~G. Lyne), from the Columbia Astrophysics Laboratory
(F. Camilo), and the author of this review.  As a result, to date a
dozen new pulsars have been discovered in 6 globular clusters, none of
which had previously known pulsars associated.  More recently, 3
further millisecond pulsars have been detected in one of these clusters
(namely NGC~6266) by Jacoby et al. (2002), exploiting the sensitivity
of the new Green Bank Radiotelescope.  So the total number of known
millisecond pulsars in globular cluster is now 57.

In the following I will focus on the globular clusters NGC~6397 and
NGC~6752, summarizing the main scientific results obtained so far,
sometimes combining radio data with optical and X-ray
observations. What I will review here have been originally reported in
seven different papers: D'Amico et al. (2001b) and Possenti et
al. (2001) described the new hardware and software systems adopted in
the Parkes search for GC-MSPs, the current status of the experiment
and its perspectives. D'Amico et al. (2001a) announced the discovery
of the first 4 millisecond pulsars detected during the Parkes search
for GC-MSPs. D'Amico et al. (2001c) and Ferraro et al. (2001)
presented the timing solution and the optical observations for the
binary millisecond pulsar in NGC~6397. D'Amico et al. (2002) and
Colpi, Possenti \& Gualandris (2002) investigated the case of
NGC~6752.

\begin{figure}
\centerline{\psfig{file=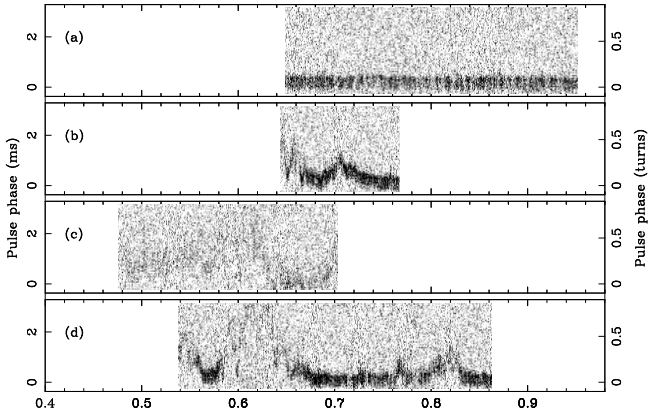,width=8.8cm,clip=} }
\caption{Observed signal intensity at 1.4 GHz as a function of
orbital phase and pulsar phase for four long observations away from the
nominal eclipse region. 
The data are processed in contiguous integrations of 120 s
duration. (a) $\sim 10$ hr observation starting on 2000 December 3 at 22:23
UT; (b) $\sim 4$ hr observation starting on 2001 March 1 at 22:32 UT; (c)
$\sim 7$ hr observation starting on 2001 March 5 at 18:31 UT; (d) $\sim 11$
hr observation starting on 2001 March 12 at 15:02 UT  
(from D'Amico et al. 2001c).
\label{image1}}
\end{figure}

\section{The millisecond pulsar in NGC~6397}

NGC 6397 is one of the most promising targets for searching MSPs in
GCs: in fact, it is one of the closest clusters, at a distance of
$2.6~{\rm kpc}~\pm 6\%$ (Reid \& Gizis 1998) and probably has a
collapsed core with hints of mass segregation (King, Sosin \& Cool
1995).  It lies in fourth place in the list of GCs ranked according to
central luminosity (Harris 1996).  In this cluster only one
millisecond pulsar has been discovered so far: PSR~J1740$-$5340, a
binary MSP with a spin period of 3.65 ms (D'Amico et al. 2001a). Its
X-ray counterpart has been identified by Grindlay et al. (2001) among
the $\sim 20$ X-ray sources detected with {\sl Chandra} within
$2\arcmin$ of the cluster center (8 of which probably being CVs).

\subsection{Observed features of PSR~J1740$-$5340}

 It displays eclipses at 1.4 GHz for more than
40\% of the orbital phase.  Ten eclipsing systems containing a MSP are
known (Fruchter et al. 1990; Lyne 1990; Stappers et al. 1996; Nice,
Arzoumanian \& Thorsett 2000; Camilo et al. 2000; D'Amico et
al. 2001a) and in one of them (PSR~B1744$-$24A: Nice \& Thorsett 1992)
the eclipses show duration and irregularities similar to those of
PSR~J1740$-$5340. However this new system is $2-3$ times wider (with
an orbital separation of $\sim 6.5~{\rm R_\odot}$) than any other
known eclipsing pulsar binary, and has a heavier minimum mass for the
companion ($>$ 0.19 M$_{\odot}$) than any known eclipsing system.

\begin{figure}
\centerline{\psfig{file=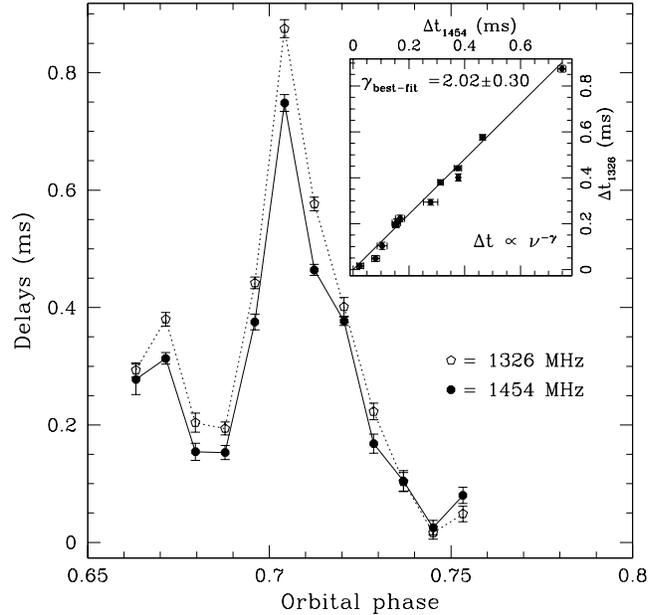,width=8.8cm,clip=} }
\caption{Excess group delays measured in two 128 MHz-wide bands
centered at 1454 MHz ({\it filled symbols\/}, connected by a {\it
solid line\/}) and at 1326 MHz ({\it open symbols\/}, connected by a
{\it dotted line\/}) for an event occurring at orbital phases
0.65$-$0.75 on 2001 March 1 (see Fig. 1b).  The delays in the two
subbands are fitted ({\it inserted panel\/}) with a straight line of
slope 1.21$\pm 0.04$, corresponding to a frequency dependence for the
delays ($\Delta t \propto \nu^{-\gamma}$) with a best-fit power-law
index $\gamma =2.02\pm 0.30$ (from D'Amico et al. 2001c).
\label{image2}}
\end{figure}

In addition, the radio signal exhibits strong fluctuations (delays
and intensity variations) over a wide range of orbital phases (see
Figure~1), indicating that the MSP is orbiting within a large envelope
of matter released from the companion with a high mass loss rate. 
The frequency-dependent behavior of the observed delay
events have been studied 
by splitting the 256-MHz bandwidth available at 1.4 GHz into
two adjacent subbands.   
Figure~2 shows an example of the measured excess delays
in the two subbands.  The small panel also shows these measurements
plotted against each other.  The best-fitting straight line
corresponds to $\Delta t \propto \nu^{-2.02\pm 0.30}$ for the excess
delays, consistent with the $\nu^{-2}$ dependency expected if the
delays are due to dispersion in an ionized medium. If entirely due to
dispersion, the excess propagation delays of up to $\sim 3~{\rm ms}$
visible in Figure~1 would correspond to electron column density
variations of $\sim 1.5\times 10^{18}~\Delta t_{-3}$ cm$^{-2}$, or
$\Delta{\rm DM}\sim 0.47~\Delta t_{-3}$ cm$^{-3}$pc ($\Delta t_{-3}$
being the delay at 1.4 GHz in ms).  This corresponds to a pulse
broadening of $0.10~P~\Delta t_{-3}=0.36\Delta~t_{-3}$ ms over the
receiver bandwidth (to be compared with an intrinsic pulse FWHM of
$0.17~P$) and may be responsible for most of the signal attenuation
and pulse broadening observed near the delay peaks.

When combined with the typical duration of the eclipse at 1.4 GHz
($\sim 13$ hr), these parameters imply an eclipse radius $R_{\rm ECL}$
($\sim 7.5$ R$_\odot$) larger than the orbital separation $a\sim 6.5$
R$_\odot$ for any value of the orbital inclination $i$. $R_{\rm ECL}$
is defined as the chord subtended on the orbital circumference of
radius $a$ by the angle between the orbital phase of eclipse ingress
(or egress) and the orbital phase 0.25 (see also Figure~5).

\begin{figure}
\centerline{\psfig{file=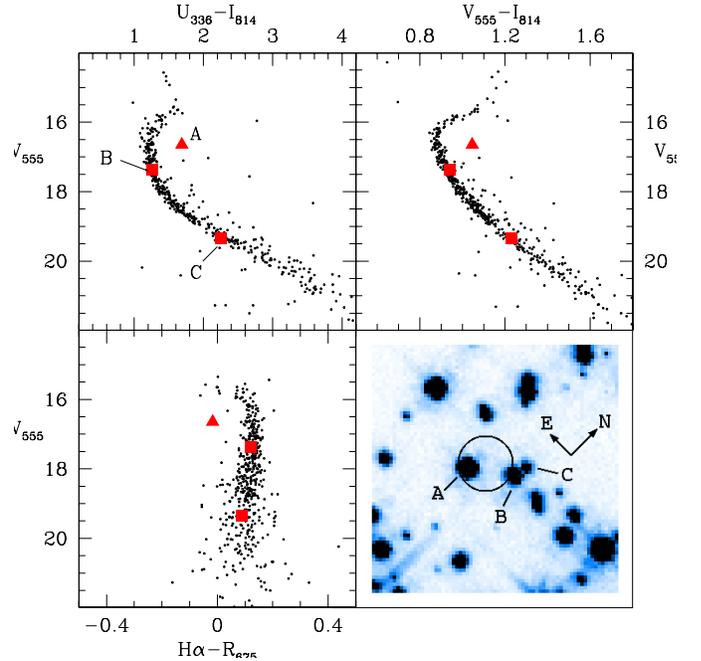,width=8.8cm,clip=} }
\caption{{\it Bottom right panel}: A portion of a median-combined
F675W HST WFPC2 image (chip WF4) of NGC 6397, centered near the MSP
position.  The region covers about $7''\times 7''$. Star A is the
companion to the MSP PSR~J1740$-$5340.  {\it Remaining three panels}:
Multiband CMDs for stars detected in a region of the WF4 chip
($40''\times40''$) containing the MSP position.  The three stars found
in the vicinity of the MSP error box are marked with different symbols
(a triangle for star A and squares for stars B and C) and labeled with
their names in the first panel (from Ferraro et al. 2001).
\label{image3}}
\end{figure}

According to different values of the orbital inclination, the values
of the Roche lobe radius $R_L$ span the interval 1.3--2.2\,R$_\odot$
(Eggleton 1983).  Because $R_{\rm ECL} > a > R_L$, the matter causing
the eclipses escapes the gravitational influence of the companion star
and therefore must be continuously replenished.  Assuming isotropic
emission, mass continuity implies that the companion star releases
mass at a rate $\dot M_{c}=4\pi R_{\rm ECL}^2\rho (R_{\rm ECL})v_f.$
Using the values at the eclipse radius inferred from the observation
of the excess delays assuming completely ionized matter ($\rho [R_{\rm
ECL}]\sim 3\times 10^{-18} \Delta t_{-3}$ g cm$^{-3}$), it results a
mass loss rate from the companion $\dot M_{c}\sim 1.5\times
10^{-11}~\Delta t_{-3}~v_{f,8}~{\rm M_\odot y^{-1}}$ where $v_{f,8}$
is the wind velocity at $R_{\rm ECL}$ in units of 10$^8$ cm s$^{-1}$
(typical order of magnitude of the escape velocity from the surface of
the companion).

The large position offset of PSR~J1740$-$5340 with respect to the
center of NGC~6397 ($r\sim 0\farcm55$) limits the unknown contribution
to the observed period derivative $\dot{P}$ due to acceleration in the
globular cluster potential. Thus $\dot{P}$ can be used to infer the
rotational energy loss of the pulsar, $\dot E\sim 1.4\times 10^{35}$
ergs s$^{-1}$.  At the distance of $6.5~{\rm R_\odot}$, the power
impinging on a white dwarf companion (of typical radius
$R_{wd}=0.033~{\rm R_\odot}$) would be $W_{wd} \la 10^{30}$ ergs
s$^{-1}$.  This would allow release of an ionized wind from the
degenerate companion at a maximum rate $\dot
M_{wd,max}=W_{wd}(R_{wd}/GM_{wd})\la 1.6\times 10^{-12}\; {\rm M_\odot
y^{-1}}.$ Thus, for realistic terminal velocity of the wind ($v_f \sim
1000$ km s$^{-1}$) the pulsar flux cannot sustain the observed
releasing of mass, $\dot M_{c},$ from a white dwarf companion.

This conclusion suggested that more likely the companion could be a
bloated main-sequence star or a sub-giant, prompting a deep optical
search for it. Using a series of public {\it Hubble Space Telescope}
archive exposures taken in 1996 March and 1999 April, Ferraro et
al. (2001) found a bright, $\sim 16.6$ mag, variable star (hereafter
Star A) with an anomalous red color and optical variability ($\sim$
0.2 mag), which nicely correlates to the orbital period of the pulsar
(see Figures 3 \& 4). This variable had been already discovered by
Taylor et al. (2001); however on the basis of the unusual colors and
time variability, Taylor et al. (2001) had classified this object as a
BY Draconis star.  Radius $R_{c}\sim 1.3-1.8~{\rm R_\odot}$ and
effective surface temperature $T_{eff}\sim 5500-5800$ K for the
companion to PSR~1740-5340 can be derived by comparing the
color-magnitude diagram in Fig.~3 with the isochrones by Silvestri et
al. (1998) and by Vanderberg (2000) for metallicity [Fe/H]=$-$2.00 and
ages of $t=12-14$ Gyr (compatible with the values observed for NGC
6397).

\subsection{The nature of PSR~J1740$-$5340}

The red color of PSR-A shows that it is not an undisturbed main
sequence star. Its peculiar nature can be unveiled inspecting the
amplitude and the shape of its light curves (insert in Figure~4).  Two
other eclipsing millisecond pulsars, both in the Galactic field, have
measured light curves: PSR B1957+20 (Callanan, van Paradijs \&
Rengelink 1995) and PSR J2051-0827 (Stappers et al.  2001).  In both
cases, the optical companions show strong modulations, which are
readily explained by the heating of one side of the companion by
radiation from the pulsar.  The optical maximum is at orbital phase
0.75, where the heated side of the companion faces the earth.  A
similar trend, though with a much smaller degree of modulation, is
seen in 47 Tuc U$_{opt}$, the first identified MSP companion in a GC
(Edmonds, et al. 2001).

\begin{figure}
\centerline{\psfig{file=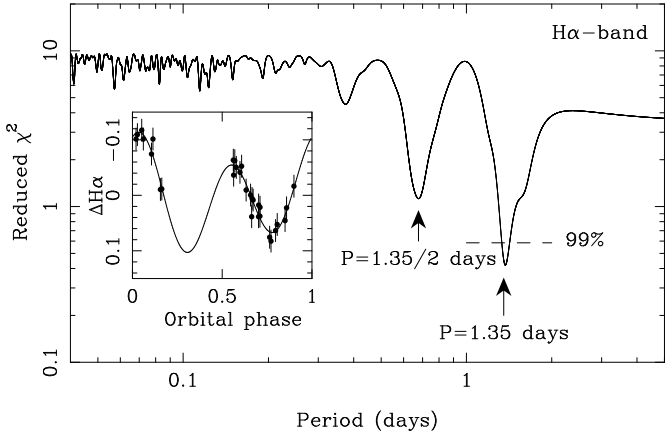,width=8.8cm,clip=} }
\caption{Reduced $\chi^{2}$ resulting from the fit of two Fourier
harmonics to the H$\alpha$ data as a function of the modulation
period. {\it Small panel}: same H$\alpha$ data folded using the period
and the reference epoch of the radio ephemeris, and best fit (solid
line) of the spectral amplitudes of the 1$^{st}$ and 2$^{nd}$
harmonics to the data (from Ferraro et al. 2001)
\label{image4}}
\end{figure}

The light curves of Star A are completely different. In Figs. 4 \& 5,
the phase 0.0 is located at the ascending node of the MSP orbit; thus
at the phase 0.75 the side of the companion facing the pulsar is
visible. In contrast to the other known variable MSP companions, the
light curves of Star A display there a minimum instead of a maximum
(see Figure~4). Within the limits in the available orbital period
coverage, the best-fit light curve of Figure~4 shows two maxima and
two minima during each binary orbit: thus, tidal distorsions appear
the more natural responsible for this pattern.  If this is the correct
interpretation, the observed modulation can be reproduced only if the
companion has almost filled up its Roche lobe and the orbital plane is
nearly edge-on ($i\sim 90^o$).

In summary, as roughly depicted in Figure~5, this binary appears the
first example of a system comprising an active MSP whose companion
keeps on pouring mass in the binary system through the inner
Lagrangian point of its Roche lobe.

\begin{figure*}
\centerline{\psfig{file=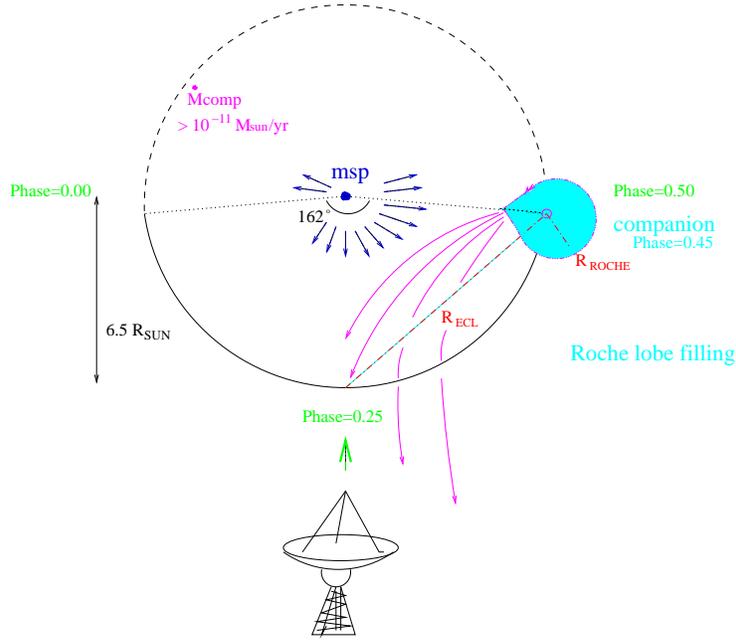,width=11.0cm,clip=} }
\caption{Sketch of the binary system containing PSR~J1740$-$5340 at
the phase of the egress of the millisecond pulsar from the eclipse at
1.4 GHz. $R_{\rm ECL}$ represents the eclipse radius, greater than the
orbital separation of the system. The maximum of the eclipse occurs at
phase 0.25 (lower side of the figure). The eclipse ingress is roughly
located at phase 0.05, whereas the eclipse egress is at phase 0.45.
\label{image5}}
\end{figure*}

\subsection{The origin of PSR~J1740$-$5340}

Three hypotheses have been suggested for explaining the origin of this
system: 
\begin{itemize}
\item[{\it (a)}] that Star A is a main sequence
(MS) star perturbed by the energetic flux emitted from the MSP; 
\item[{\it (b)}] that PSR~J1740$-$5340 is a new-born MSP, the first
one observed just after the end of the process of recycling (D'Amico
et al. 2001c; Burderi, D'Antona \& Burgay 2002; Ergma \& Sarna 2002);
\item[{\it (c)}] that PSR~J1740$-$5340 is
a re-born MSP, which has already experienced many phases of recycling
(Grindlay et al. 2002).
\end{itemize}

Given the relatively large orbital separation, it seems energetically
difficult to explain an increasing of both the effective temperature
and of the radius of a main sequence companion: however further
detailed simulations are required before excluding the scenario {\it
(a)}.  The hypothesis {\it (b)} is particularly stimulating because
could be the first example of a binary in the so-called {\it
radio-ejection} phase (Burderi, D'Antona \& Burgay 2002) in which the
Roche-lobe overflow stage is not ended yet, but the MSP is already
active and sweeps away the infalling matter preventing accretion onto
its surface.  In particular (Burderi, D'Antona \& Burgay 2002) Star A
could have been originally a main sequence star of $\sim 1~{\rm
M_\odot}$, whose evolution triggered mass transfer towards the compact
companion, spinning it up to millisecond periods (Alpar, et
al. 1982). Irregularities in the mass transfer rate from the
companion, $\dot M_{c},$ are common in the evolution of these systems
(e.g. Tauris \& Savonije 1999): even a short decreasing of $\dot
M_{c}$ can have easily allowed PSR~J1740$-$5340 (having a magnetic
field $\sim 8\times 10^8$ G and a rotational period $\la 3.65$ ms) to
become source of relativistic particles and magnetodipole emission,
whose pressure {\it (i)} first swept the environment of the "neutron
stars (NS), allowing coherent radio emission to be switched on
(Shvartsman 1970) and {\it (ii)} then kept on expelling the matter
overflowing from the Roche lobe of Star A (Ruderman, Shaham \& Tavani
1989). For a wide enough binary system, once the radio pulsar has been
switched on, any subsequent restoration of the original $\dot M_{c}$
cannot quench the radio emission (Burderi et al. 2001). Remarkably,
for the parameters of PSR~J1740$-$5340, the orbital period at which
{\it radio-ejection} dominates is $\sim 39$ hrs, very close to the
observed value of $\sim 32.5$ hrs (Burderi, D'Antona \& Burgay 2002).

As already noticed, the signal from PSR J1740$-$5340 is eclipsed for
about 40\% of the time at 1.4 GHz and strongly disturbed along all the
orbit (D'Amico et al. 2001c) and this suggests a bias against the
detection of this kind of radio sources.  On the other hand, the
occurrence of the {\it radio-ejection} mechanism can strongly shorten
the duration of the bright X-ray phase in the recycling
scenario. Hence, if this {\it radio-ejection} phase is common in the
formation of the MSPs, the birth-rate of the Low-Mass X-ray Binaries
(depending on their observed number in the Galaxy and on the supposed
lifetime) should be revised upwards.  This could help solving the
apparent slight discrepancy between the observed birth-rate of the
MSPs and that of the Low-Mass X-ray Binaries, which are their supposed
progenitors.

Finally, hypothesis {\it (c)} calls for the high rate of dynamical
interactions between single stars and binaries allowed by the crowded
environment of the central regions of a globular cluster. Due to such
interactions, a neutron star in a globular cluster could undergo many
phases of {\it recycling}, accreting mass from different companion
stars acquired at different times along the cluster evolution. These
repeated stages of accretion could affect e.g. the evolution of the
surface magnetic field of the MSPs in globular cluster and their X-ray
luminosity (Grindlay et al. 2002).
 
\section{The five millisecond pulsars in NGC~6752}

NGC~6752 is one of the globular clusters with the more precise
distance measurement: $4.1~{\rm kpc}\pm 5\%$, obtained fitting the
white dwarf (WD) sequence (Renzini et al. 1996).  It is classified as
a core-collapsed cluster with evidence of mass segregation (Ferraro et
al. 1997). Rubenstein \& Bailyn (1997) estimated a binary fraction for
main sequence stars in the range 15\%$-$38\% inside the inner core
radius ($< 11''$), decreasing to less than 16\% beyond that. At least
6 dim X-ray sources had been identified on the basis of ROSAT
pointings (Verbunt \& Johnston 2000), whereas a more recent {\it
Chandra} observation revealed 19 sources within the half-mass radius
of the cluster (Pooley et al. 2002).

In this cluster a 3.26 ms binary pulsar was first discovered, and
originally labeled PSR~J1910$-$59A (D'Amico et al. 2001a).  Amplification due
to scintillation helped in the detection of four additional MSPs in
the same cluster (hereafter PSRs B, C, D, E).  All of them are
isolated with spin periods in the range $4.6-9.0$ ms.  X-ray
counterparts for PSR$-$D, PSR$-$C and PSR$-$B have been detected
(Pooley et al. 2002, D'Amico et al. 2002) using a 28700 sec long
exposure taken with the ACIS-S detector aboard the {\it Chandra} X-ray
observatory.

\subsection{The three inner millisecond pulsars}

PSRs B, D and E are located close to the cluster center, as expected
as a consequence of mass segregation in the cluster (see Figure~6).
In particular, PSR$-$D has the third largest period derivative,
$\dot{P} = 9.6\times 10^{-19},$ among known MSPs (after PSR
B1820$-$30A in NGC~6624 and PSR B1821$-$24 in M28), suggesting that
the $\dot{P}$ is dominated by the line-of-sight acceleration in the
cluster gravitational field. This interpretation is supported by the
large negative $\dot{P}$ values observed for PSR$-$B and E, which
allow to directly derive lower limits to the line-of-sight
accelerations, $a_{l}/c=\dot{P}/P = -9.6\pm 0.1\times 10^{-17}$
s$^{-1}.$ This value is the largest known after those of
PSRs~B2127+11A and D in the core of M15 (Anderson et al. 1990).  Other
contributions to $a_{l}/c$ are all negligible (D'Amico et al. 2002)
and thus the inferred high values of $|a_{l}/c|$ unambigously calls
for the effects of the potential well of NGC~6752.

\begin{figure}
\centerline{\psfig{file=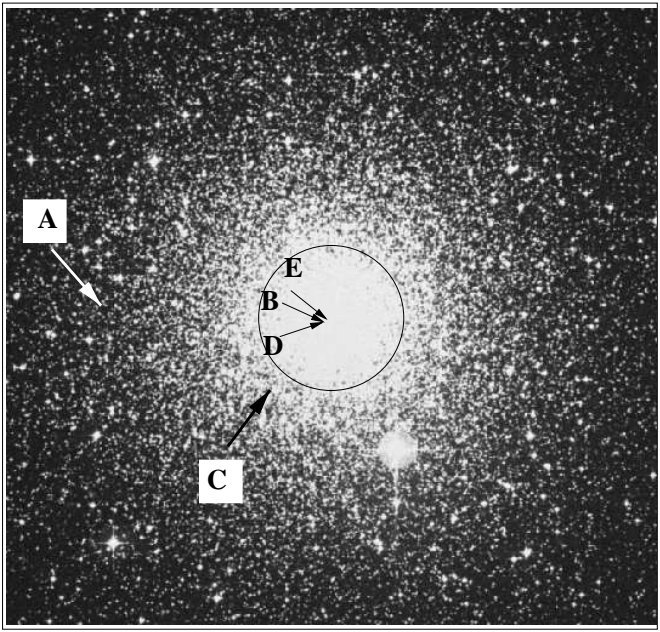,width=8.8cm,clip=} }
\caption{Positions of the 5 millisecond pulsars (arrows) in NGC~6752,
superimposed on an optical image retrieved from the Digital Sky
Survey. The circle indicates the half-mass radius region
($\theta_{hm}=115\arcsec$).
\label{image6}}
\end{figure}

\subsection{Central Mass-to-Light ratio in NGC~6752}

A lower limit to the mass-to-light ratio in the inner regions
of NGC~6752 can be derived from the following rule, which holds to
within $\sim$10\% in all plausible cluster models (Phinney 1992):
\begin{eqnarray*}
 & \left|\frac{\dot{P}}{P}(\theta_{\perp})\right| <
\left|\frac{a_{l,max}(\theta_{\perp})}{c}\right| \simeq & \\
 & \simeq 1.1\frac{G}{c}\frac{M_{cyl}(<\theta_{\perp})}{\pi
D^2\theta^2_{\perp}}=5.1\times 10^{-18} \frac{\cal{M}}{{\cal
L}_V}\left(\frac{\Sigma_{V}(<\theta_{\perp})}{10^4~{\rm L_{V\odot}
pc^{-2}}}\right){\rm s^{-1}}.
\label{eq:Sigma}
\end{eqnarray*}
Here $\Sigma_{V}(<\theta_{\perp})$ is the mean surface brightness
within a line of sight subtended by an angle $\theta_{\perp}$ with
respect to the cluster center, $M_{cyl}(<\theta_{\perp})$ is the mass
enclosed in the cylindrical volume of radius
$R_{\perp}=D\theta_{\perp}$ and $\cal{M}/{\cal L}_V$ is the mean {\it
projected} mass-to-light ratio in the V-band.  In Figure~7 the curves
of maximum $|a_l/c|$ for different values of $\cal{M}/{\cal L}_V$ have
been plotted, using the most recent published brightness profile for
this cluster (Lugger, Cohn \& Grindlay 1995), normalized to the
central surface brightness in the V-band reported by Djorgovski
(1993). In particular, in Figure~7 the histogram represents the data
for $\cal{M}/{\cal L}_V$=1.1, the mass-to-light ratio suggested by
Pryor \& Meylan (1993) for the central regions of NGC~6752. This
greatly underestimates the observed values of $\dot{P}/P$.  The dashed
lines are good analytical fits to the observed data, scaled according
to increasing values of $\cal{M}/{\cal L}_V$. Only $\cal{M}/{\cal
L}_V\gapp$9 can account for the observed $|\dot{P}/P|$ of PSRs B and
E. An even larger $\cal{M}/{\cal L}_V\gapp$ 13 is required if PSR-D
has a negligible intrinsic $\dot{P}/P$ (see caption of Figure~7).  It
is worthwhile noting that these estimates of the projected
$\cal{M}/{\cal L}_V$ are independent of modeling of the cluster
potential and of distance, excepting the effects of extinction.  Being
E(B$-$V) very small for NGC~6752 (= 0.04 according to Harris 1996) the
latter affects negligibly the result.

It provide the first direct dynamical evidence for a high density of
unseen remnants in the core of a globular cluster, suggesting a
mass-to-light ratio $\gapp$10 in the core region.  Typical
mass-to-light ratios for the central regions of the globular cluster
span the interval $\cal{M}/{\cal L}_V$=1$-$3.5 (Pryor \& Meylan 1993).
NGC~6256, the GCs at the top of a list ordered
according to ${\cal M}/{\cal L}_V$ (Pryor \& Meylan 1993), has ${\cal
M}/{\cal L}_V\sim 6$, but this value is quite uncertain due to
the high level of reddening. Using the same method applied to
NGC~6752, a much smaller value $\sim 3,$ was obtained by Phinney
(1993) in the case of M15, a core-collapsed GC long suspected to host
a central black-hole (see e.g. van der Marel \& Roeland 1999).

The nature of the remnants in the core of NGC~6752 is not clear as it
strongly depends on assumptions about the initial mass function of the
cluster.  However, the possibility that they are black holes, or that
many stellar remnants have collapsed into a massive (binary?) black
hole, is intriguing.

\subsection{The millisecond pulsar in the outskirts of the cluster}

When a coherent timing solution became available, PSR$-$A (the only
binary pulsar known in the cluster) resulted located at $6\farcm4$
from the globular cluster center (R.A.=19$^{\rm h}$ 10$^{\rm m}$
$51\fs8,$ Dec.=$-$59$^\circ$ 58$\arcmin$ 55$\arcsec,$ eqx 2000, Harris
1996) equivalent to 3.3 half-mass radii or $\sim 57$ core
radii. Hence, it is the farthest objects from the GC center known in
the catalogue of 41 pulsars in GC with accurately determined
positions.  Previously, the largest offset of an associated pulsar
from a GC center was for PSR B2127+11C (Prince et al. 1991), a member
of a double-neutron-star eccentric binary in M15, located at $\sim$
0.8 half-mass radii.  Given this large radial offset, one could
question the association of PSR$-$A with the cluster.  However, the
probability of chance superposition of a Galactic field MSP with a
flux density of $S_{1400}\gapp~0.2$ mJy within $6\farcm4$ of the
center of NGC~6752 is at most $10^{-5}$.  This probability is further
reduced by noting that all five pulsars have very similar DM values.

The companion to PSR$-$A is a dwarf (probably degenerate) star of mass
$\mdeg\sim 0.2\msun$, moving with period $\porb=0.86$ days on a
circular orbit around PSR$-$A (the eccentricity is $<10^{-5}$, D'Amico
at al. 2002). The orbital separation is $\apul\sim 0.0223$ AU, and the
binary has a gravitational binding energy $E_{\rm PSR-A}\sim
10^{47}\erg.$ The inferred magnetic field is $\sim 10^8$ Gauss and the
characteristic age $\tau_{\rm {PSR-A}}\sim 15$ Gyr (given the large
offset from the GC center, the period derivative of PSR$-$A is not
significantly affected by the GC potential well); in short, it looks
like a "canonical" recycled MSP (Lorimer 2001).

\begin{figure}
\centerline{\psfig{file=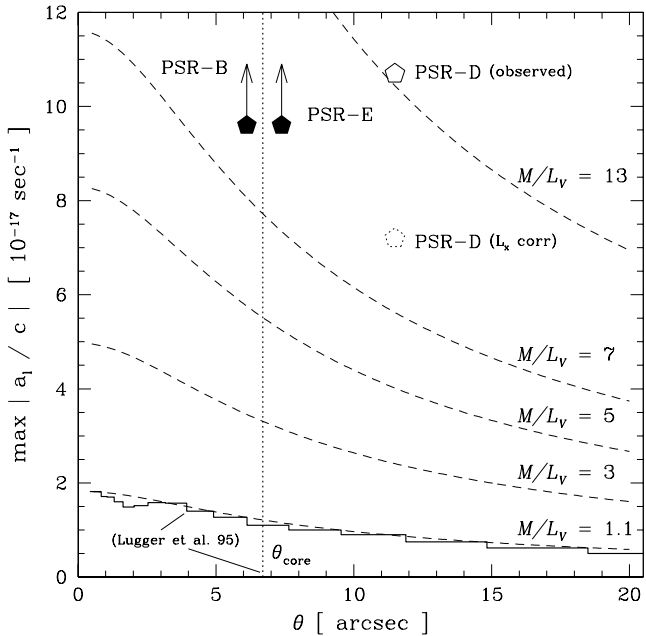,width=8.8cm,clip=} }
\caption{Maximum line-of-sight acceleration $|a_{{l}_{max}}/c| =
|\dot{P}/P|$ versus displacement $\theta$ with respect to the center
of NGC~6752.  The histogram represents the prediction based on the
available optical observations. The dashed lines are analytical fits
to the optical observations, labeled according to the adopted
mass-to-light ratio. The filled pentagons represent lower limits to
the line-of-sight acceleration based on the observation of PSRs B and
E. The open pentagon shows the value of $\dot{P}/P$ for PSR-D assuming
a negligible intrinsic $\dot{P}$.  If various observed scalings
(Becker \& Tr\"umper 1997; Possenti et al. 2002; Grindlay et al. 2002)
between X-ray luminosity and spin-down power for the MSPs are used to
estimate the intrinsic $\dot{P}$, $\dot{P}/P$ for PSR-D assumes values
in the interval between the open and the dotted pentagon (from D'Amico
et al. 2002).
\label{image7}}
\end{figure}

\subsection{The origin of PSR~J1911-5958A}

Colpi, Possenti \& Gualandris (2002) explored three possible
explanations for the unprecedented position of PSR~J1911$-$5958A (this
is the correct name given to the pulsar after the determination of its
precise position):
\begin{itemize}
\item[{\it (a)}] that PSR$-$A, remnant of a massive star, was the
component of a truly {\it primordial binary} born in the GC halo or
expelled from the GC core due to the natal kick imparted by the
supernova explosion;
\item[{\it (b)}] that it was ejected from the core via {\it scattering}
or {\it exchange} interaction(s) with cluster {\it stars};
\item[{\it (c)}] that it was propelled into the cluster halo by a {\it
scattering} event involving a {\it black hole binary} (hereafter [BH+BH]). 
\end{itemize}

\subsubsection{Case {\it (a)}}

Hypothesis {\it (a)} is the simplest among the three. It requires
orbit stability against dynamical friction (having a characteristic
time scale $\tau_{\df}$), and a low recoil speed at the moment of the
neutron star (NS) formation.

If the binary has been kicked from the globular cluster central
regions into a highly eccentric orbit ($e_{\rm ec}\simgreat 0.9$) at
the time of NS formation, it would remain at the periphery for a time
$\tau_{\df}\simgreat 7\times 10^8 (1.6\msun/M)$ yr before returning to
the core. Hence, if the supernova took place in the central regions,
the binary of PSR$-$A (of mass $\simgreat 2.5\msun$ after the NS has
formed) would have been driven back to the core by dynamical friction
within $\simless 1$ Gyr.  As the supernova must have occurred in the
first Gyr since GC formation, the binary would not have survived in
the halo until present time.

For a centrophobic halo orbit ($e_{\rm ec}\simless 0.5$) the dynamical
friction timescale $\tau_{\df}$ is $\simgreat 10^{11} ({1.6\msun}/M)$
yr in NGC~6752, implying stability of the orbit of PSR$-$A and of its
progenitor halo binary.  However, if PSR$-$A indeed descends from a
binary born in the cluster outskirts and following an almost circular
orbit in the cluster potential, it must have experienced the faintest
supernova kick ever observed: in fact the escape velocity $V_{\rm
esc}$ is only $\sim 10 \kms$ in the halo of NGC~6752. Conversely, it
is known from current proper motion measurements of 13 MSPs belonging
to the Galaxy (Toscano et al. 1998) and from statistical studies
(Cordes \& Chernoff 1997) that MSPs have relatively large mean
transverse velocities of 85$\pm13\kms$ implying peculiar speeds with a
3-D dispersion of $84~\kms.$ Moreover, theoretical studies estimates
that the minimum recoil speeds $V_{\rm rec}$ for a MSP binary is
$\simgreat 60\kms$ (Tauris \& Bailes 1996).  As a consequence,
scenario {\it (a)} calls for the NS of PSR$-$A being probably formed
from the accretion-induced collapse of a massive white dwarf (Grindlay
et al. 1988), in which a large kick velocity is not expected.

A difficulty of this scenario is related with the probability that a
MSP is formed in the outer halo of a globular cluster. It contains
only a small fraction of the total mass of the cluster and the
relatively small stellar density does not permit the occurrence of a
significant number of dynamical interactions. Hence the only available
channel for the formation of a MSP is just the evolution of a
primordial binary, similar to the formation scenario holding in the
Galactic field.  Given the estimated number $N_{\rm gal}\sim4\times
10^4$ of MSPs in the Galactic field (e.g. Lorimer 2001) and the ratio
$\mu\simless 5\times 10^{-4}$ between the mass in all the globular
clusters to that of the Galaxy, it turns out that at most $N_{\rm
gcs}=N_{\rm gal}\mu f_r\sim 20f_r$ MSPs born from a primordial binary
should be hosted by the GCs system (comprising $\sim 200$ objects) at
radii larger than $r$. Here $f_r$ is the fraction of the total mass in
a typical GC which resides at a distance greater than $r$ from the
center. Taking $r$ equal to 3 half-mass radii, $f_r\sim 0.15$ and thus
$N_{\rm gcs}\sim 3$.  The very low number $N_{\rm gcs}$ makes
improbable the detection of a MSP at its offset position, when
descending from a primordial binary born in the GC halo.

\subsubsection{Case {\it (b)}}

Ejection of PSR$-$A away of the core {\it via} a dynamical interaction
with a cluster {\it star} calls for two possible scenarios: {\it
(b1)} inelastic scattering or {\it (b2)} three-body exchange.
Scattering can occur either before or after recycling as it preserves,
to a high degree, the circularity seen in our binary pulsar. In both
cases, a recoil kinetic energy of $2-3\times 10^{46}\erg$ is requested
for ejecting [NS$_{\rm A}$+co] (comprising either PSR$-$A or the not
yet recycled NS and its companion) into its halo orbit.  This energy
must be extracted from the [NS$_{\rm{A}}$+co] binding energy:
inspecting the tables of Sigurdsson \& Phinney (1993) it is easy to
conclude that this energetic requirement imposes an upper limit on the
orbital period $P_{\rm {orb,up}}$ of [NS$_{\rm{A}}$+co] which is
$\simless 0.4$ days.

At the onset of recycling a "bifurcation" orbital period, $P_{\rm
bif}\sim 2$ days, separates the formation of a converging binary
(which evolves with decreasing orbital period $P_{\rm orb}$ until the
mass-losing star becomes degenerate and a short-period binary pulsar
is formed) from a diverging one (which evolves with increasing $P_{\rm
orb}$ forming a long period binary pulsar; Tauris \& Savonije 1999).
Given this scenario, PSR$-$A should result from a converging binary
and thus its period can not have ever been shorter than its current
value of 0.86 days and this exclude the scattering hypothesis {\it
(b1)}.

Exchange (scenario {\it (b2)}) alters the nature of the interacting
system and makes the new binary rather eccentric and with a longer
orbital period. Thus due to the very long circularization time between
compact stars, a three-body exchange must occur before recycling
(reminding that the upper limit on the eccentricity of the orbit of
PSR$-$A is 10$^{-5}$).  The simplest and most probable case in this
scenario is that of a main sequence binary [MS+ms] (MS heavier than
ms) interacting with the NS$_{\rm A}$ (slowly rotating not yet
recycled progenitor of PSR$-$A) flying by (given the similar masses of
MS and NS$_{\rm A}$, the results are almost unaffected when swapping
their roles).  In such a three-body encounter the energetic
requirement on the recoil kinetic energy of [NS$_{\rm{A}}$+co] can be
accomplished, provided that the original main sequence binary is tight
enough. On the other hand the tables of Sigurdsson \& Phinney (1993)
show that after the exchange, the [NS$_{\rm A}$+co] binary has an
orbital period suspiciously close to the currently observed value,
considering that recycling (and hence shrinking of the orbit) has not
occurred yet. Even more disturbing is the value of the characteristic
time ($\sim 18$ Gyr $>$ age of the cluster) for an interaction of this
kind to take place in the core of NGC~6752.

\subsubsection{Case {\it (c)}}

Partly motivated by the unusual high mass-to-light ratio discovered in
the inner region of NGC~6752 (see previous section), Colpi, Possenti
\& Gualandris (2002) investigated a more exotic scenario, involving
the presence of a binary black-hole in the centre of the cluster.  In
effect, dynamical studies (Portegeis Zwartz \& McMillan 2000) show
that some clusters could still retain one binary, the last which has
not been kicked out.  This would be the consequence of repeated
exchange interactions and ejections (Sigurdsson \& Hernquist 1993;
Portegeis Zwartz \& McMillan 2000), occurring early in the cluster
lifetime among the initially rich population (of few hundreds) of BHs,
born from the most massive stars in the cluster. As they sink toward
the center by dynamical friction (on a timescale $\simless 10^6$ yr),
BH$-$[BH+BH] interactions become overwhelmingly important. Binaries
hardens progressively in virtue of these encounters and eventually
leave the cluster due to recoil. As a result, many clusters lose their
original BHs on a time scale of $\sim$ Gyr. But the outcome of the
last encounter (which could involve two single BHs or a single BH and
a binary BH system) depends on the specific and unpredictable
parameters of the interacting corpses and thus could either eject all
the 3 bodies or leave a single BH or leave a binary BH.  Following
Miller \& Hamilton (2002a), the supposed binary that avoided ejection
should likely have unequal masses and could harbor the heaviest BH
that NGC~6752 ever had.

\begin{figure}
\centerline{\psfig{file=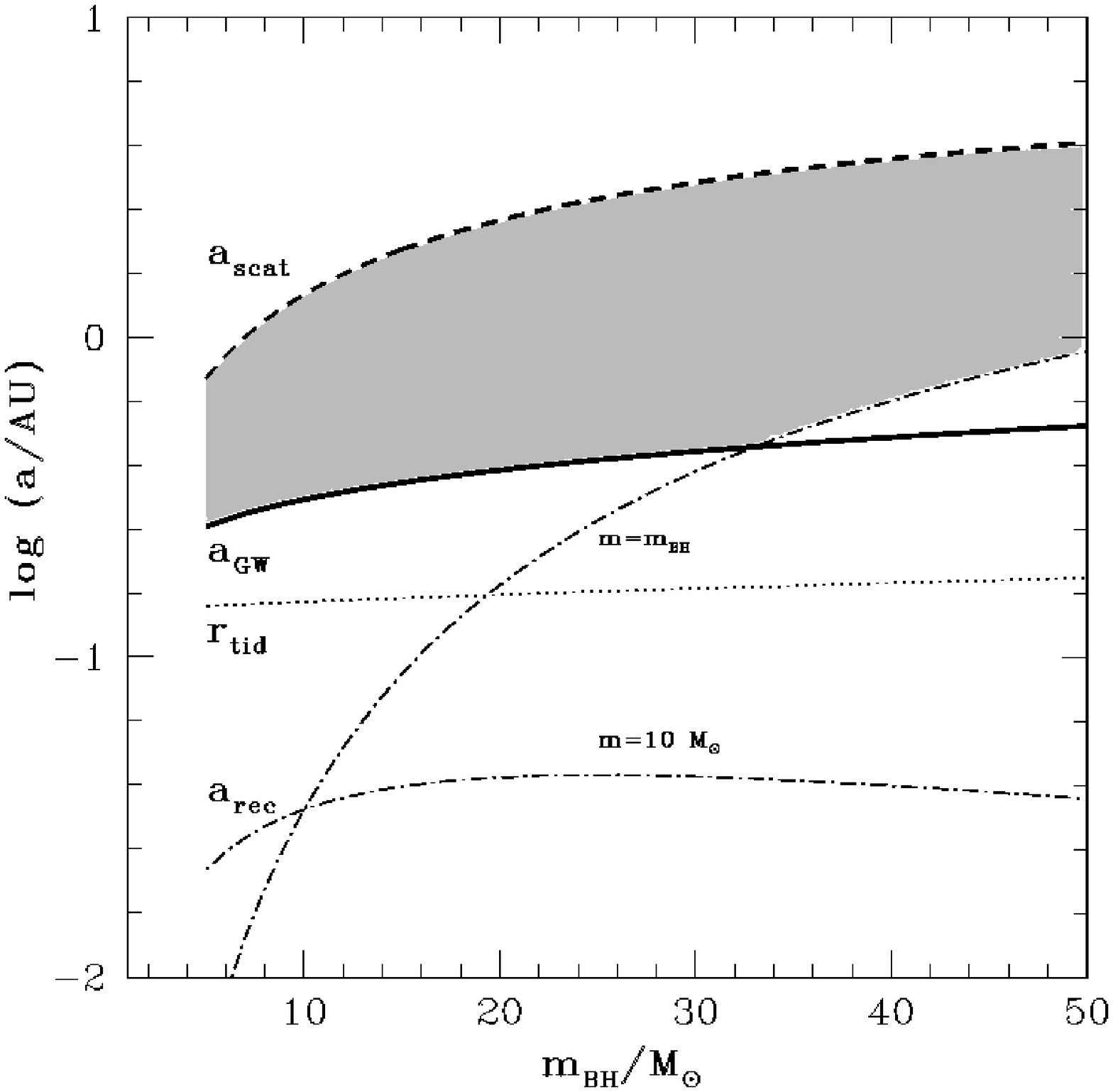,width=8.8cm,clip=} }
\caption{[BH+BH] binary separation $a$ against $\mbh2/\msun$ for
$\mbhbig=50\msun;$ $a_{\rm GW}$ ({\it heavy solid line}), $a_{\rm
{rec}}$ ({\it heavy dot-dashed lines}) and $a_{\rm{scat}}$ ({\it heavy
dashed line}) are computed from the related equations presented in the
paragraph \ref{eq:a_rec}. In particular $a_{\rm {rec}},$ is computed
for the case of light $m=10\msun$ black-hole intruders and for the
case of intruders as massive as the secondary black-hole, namely
$m=\mbh2.$ The {\it dotted line} gives the tidal disruption radius for
[NS$_{\rm A}$+co]. The shaded area indicates the values of $a$ for
ejection of PSR$-$A during the last Gyr, compatible with its detection
(from Colpi, Possenti \& Gualandris 2002).
\label{image8}}
\end{figure}

The scenario {\it (c)} is fully consistent and viable provided that:
\begin{itemize}
\item[{\it i}] the [BH+BH] binary that survived the early ejection
from the GC, later avoided also coalescence by emission of
gravitational waves (GWs), at least for a time $\tau_{\rm {GW}}\ge 10$
Gyr;
\item[{\it ii}] the separation of the [BH+BH] binary is sufficiently
large to have prevented ejection from the cluster due to a single
interaction with a much lighter stellar intruder;
\item[{\it iii}] the hardening of the [BH+BH] binary due to the
encounters with the bath of the cluster stars impinging on it does not
reduce its orbital separation enough for triggering coalescence by
emission of GWs (point {\it i}) or ejection (point {\it ii});
\item[{\it iv}] during the scattering interaction, the [NS$_{\rm
A}$+co] binary is not ionized by the [BH+BH] binary tidal field;
\item[{\it v}] the [BH+BH] binary can have transferred enough kinetic
energy to propel PSR$-$A into the cluster halo;
\item[{\it vi}] both the probability of occurrence of the scenario
{\it (c)} and the probability of detection of PSR$-$A at the present
offset position are not negligible.
\end{itemize}

Indicating with $\mbhbig$ and $\mbh2$ the masses of the two BHs in the
binary, with $\bhtot$ and $\mubh$ their total and
reduced mass, and with $m$ the mass of an intruder, 
a [BH+BH] binary (with typical eccentricity of $0.7$)
satisfies the condition {\it i)} if its separation $a$ exceeds a
critical value $a_{\rm GW},$ that is
$$
a > a_{\rm {GW}}\sim 0.4 \left
({\bhtot \over 100\msun}\right )^{1/2} \left ({\mubh \over
10\msun}{\tau_{\rm {GW}}\over 10^{10} \rm {yr}}\right )^{1/4}\!\!\!
\rm {AU}~.
\label{eq:a_gw}
$$
On the contrary the condition {\it ii)} imposes the inequality 
$$
a > a_{\rm {rec}}\sim
0.01 \left ({\mubh\over 10 \msun} \right)\!\!\!
\left ({m \over 10\msun} {100\msun\over \bhtot}
{35\kms \over V_{\rm{esc}}} \right)^2\!\!\!\rm {AU}~.
\label{eq:a_rec} 
$$
For the parameters of NGC~6752, the 
(very approximate) number of stars impinging onto the target
binary over a time $t\sim 10^{10}$ yr is 
$$
{\cal {{N}}_*} \simless 2\cdot 10^3~, 
\label{eq:n_dot}
$$
whereas the minimum required number of encounters for triggering the 
coalescence of the [BH+BH] binary due to gravitational waves emission is
$$
{\cal {N}}_{\rm {GW}}\sim 5\cdot 10^{3}~.
\label{eq:n_gw}
$$
Hence condition {\it iii)} is accomplished.
\noindent
The condition {\it iv)} requires that the distance of closest approach between
the two binaries satisfies the relation
$$
r_{\rm min} \sim a > r_{\rm tid} \simeq 0.2 \left({\bhtot\over 100
\msun}{1.6\msun\over \mpul+m_{\rm c}}\right )^{1/3} \!\!\!\!\!{\rm
{AU}}~.
\label{eq:r_tid}
$$
The condition {\it v)} on the conversion of the binding 
energy of the BH binary into kinetic energy for the ejection 
(with initial velocity $V_{\rm{ej}}$) of 
the [NS$_{\rm A}$+co] binary can be accomplished if
$$
a < a_{\rm{scat}}\sim \left ( {G\mubh\over V^2_{\rm{ej}}}\right ) 
\sim 1.3 \left({\mubh\over 10\msun} \right )
\left ({35\kms\over V_{\rm{ej}}}\right)^{2}\rm {AU}~.
\label{eq:a_scat}
$$ Finally, one can estimate a rate for MSP ejection ${\cal{R}}_{MSP}$
of $\sim 0.5~{\rm Gyr}^{-1},$ implying a probability of detection of
about $\sim 50\%$ considering a life-time for the current orbit of
$\sim 1$ Gyr.  The value of ${\cal{R}}_{MSP}$ can be derived on the
basis of the flux of stars on the [BH+BH] over the cluster lifetime
(see the equation for $\cal{N}_*$), assuming that about 1/30 of
these encounters should have involved a NS and that $\sim 5\%$ of the
NSs is in the state of a MSP. These figures are calculated adopting a
Salpeter-like initial mass function (giving $\sim 1\%$ of NSs) and a
mass segregation enhancement of $\sim 6$ in the density (Sigurdsson \&
Phinney 1995). If the IMF is flatter (as suggested by the observations
of D'Amico et al. 2002), this rate could be higher, enhancing the
likelihood of this scenario.  Once ejected into an halo orbit,
[NS$_{\rm A}$+co] spends only $\sim 1/50$ of its life in the dense GC
core and thus has negligible probability to undergo further
interactions.

Figure~8 summarizes the results of all the previous considerations,
showing that a dynamical interaction with a [BH+BH] binary in
a mass range of $\sim 3-100\msun$ and separation of $\sim$ 0.3-2 AU
can propel PSR$-$A into the cluster halo, being consistent with
the conservation both of energy and linear momentum. In particular
an unequal mass binary with a heavy $\simgreat 100\msun$ BH and a
lighter companion $\sim 10\msun$ is preferred, as it can form in a GC
(Miller \& Hamilton 2002b) and can provide the right statistics for
explaining PSR$-$A. 

\section{Conclusion}

During the last two years, the detection of 15 millisecond pulsars 
in six globular clusters which contained no previously known pulsars
(Possenti et al. 2001; Jacoby et al. 2002)
broke a 7 years long hiatus in such discoveries. 
More recently, the first stage of timing observations for most of the new 
objects has been completed, opening the possibility of 
investigating both their properties and the characteristics of the
globular clusters in which they are embedded. 
Here I have presented the first results of these studies in the cases of
the globular clusters NGC~6367 and NGC~6752. They confirm that
the efforts devoted to the search and timing 
of millisecond pulsars in globular clusters, when combined with
observation at other wavelengths, can deserve a significant and often
surprising scientific pay-back.

\begin{acknowledgements}
We gratefully acknowledge the support by the Heraeus foundation.
\end{acknowledgements}



\clearpage

\end{document}